\documentclass{svjour3}                     
\usepackage{hyperref}
\usepackage[utf8]{inputenc}
\usepackage{amsmath} 
\usepackage{graphicx}
\usepackage{pgfplots}   
\pgfplotsset{compat=1.13}

\newcommand{\bw}{\mathcal{W}}                   

\journalname{Foundations of Physics}
\begin{document}

\title{Decoherence and determinism in a one-dimensional cloud-chamber model}


\author{{Jean-Marc Sparenberg \and David Gaspard}}


\institute{Jean-Marc Sparenberg \href{mailto:jmspar@ulb.ac.be}{jmspar@ulb.ac.be}
\href{http://orcid.org/0000-0002-4154-5766}{ORCID 0000-0002-4154-5766} \\
David Gaspard \href{http://orcid.org/0000-0002-4449-8782}{ORCID 0000-0002-4449-8782}
\at Nuclear Physics and Quantum Physics, École polytechnique de Bruxelles,
Université libre de Bruxelles (ULB), CP 229, B-1050 Brussels, Belgium
}

\date{Received: 7 February 2017 / Accepted: 13 March 2018 \\
\href{https://doi.org/10.1007/s10701-018-0155-2}{https://doi.org/10.1007/s10701-018-0155-2}}

\maketitle

\begin{abstract}
The hypothesis \cite{sparenberg:13} that the particular linear tracks
appearing in the measurement of a spherically-emitting radioactive source in a cloud chamber
are determined by the (random) positions of atoms or molecules inside the chamber
is further explored in the framework of a recently established one-dimensional model \cite{carlone:15}.
In this model, meshes of localized spins 1/2 play the role of the cloud-chamber atoms
and the spherical wave is replaced by a linear superposition of two wave packets moving from the origin to the left and to the right,
evolving deterministically according to the Schrödinger equation.
We first revisit these results using a time-dependent approach,
where the wave packets impinge on a symmetric two-sided detector.
We discuss the evolution of the wave function in the configuration space
and stress the interest of a non-symmetric detector in a quantum-measurement perspective.
Next we use a time-independent approach to study the scattering of a plane wave on a single-sided detector.
Preliminary results are obtained, analytically for the single-spin case and numerically for up to 8 spins.
They show that the spin-excitation probabilities are sometimes very sensitive to the parameters of the model,
which corroborates the idea that the measurement result could be determined by the atom positions.
The possible origin of decoherence and entropy increase in future models is finally discussed.
\end{abstract}

\section{Introduction}

Since the birth of quantum mechanics, the question of determinism has been central.
While all previous science relied on determinism,
in particular through the concept of reproducibility (the same causes always produce the same effects),
quantum mechanics broke with this concept through the irreducible appearance of probabilities in its postulates
(the same causes do not always produce the same effects).
In particular, this happens in a measurement process
where identical measurements on identical systems in identical states may lead to different results.
More generally, each time decoherence takes place,
a linear superposition of states seems to randomly and irreversibly decohere to a statistical mixture of states.

These phenomena always happen when the quantum system interacts with a macroscopic system (e.g.\ a readable apparatus)
which displays uncontrollable internal or external (``environment'') degrees of freedom.
A natural explanation of the apparent randomness could then be the very existence of these degrees of freedom,
which would play the role of ``hidden variables'' characterizing the apparatus\footnote{
The present apparatus hidden variables are thus standard physical quantities.
They are ``hidden'', in the sense that they are not easily accessible experimentally,
but they are not ``hidden variables'' of the kind postulated by Einstein, Podolsky and Rosen \cite{einstein:1935},
which would have a different -- and unknown -- nature.}.
It was shown in reference \cite{sparenberg:13} that the fact that these variables characterize the apparatus
leads to a violation of Bell's inequalities,
in contrast with usual hidden variables which characterize the microscopic quantum system.
The theory thus agrees with standard quantum mechanics,
with a strong non-locality potentially enclosed in the quantum-system wave function.

A striking test model for apparatus hidden variables is the so-called Mott problem \cite{mott:1929,mott:95},
i.e.\ the detection of an $\alpha$-particle emitted by a radioactive source in a cloud chamber (or any modern type of track chamber).
While the quantum description of the emission is a spherical wave,
the actually detected tracks are linear (but randomly distributed with an isotropic probability distribution).
In an apparatus-hidden-variable approach,
the direction of a particular track would be determined by the positions of the detector-gas atoms or molecules.
These move randomly and can be considered as fixed for a particular emission,
given the different orders of magnitude of the $\alpha$-particle and gas-atom energies.

\begin{figure}[ht]
\centering
\includegraphics[scale=0.65]{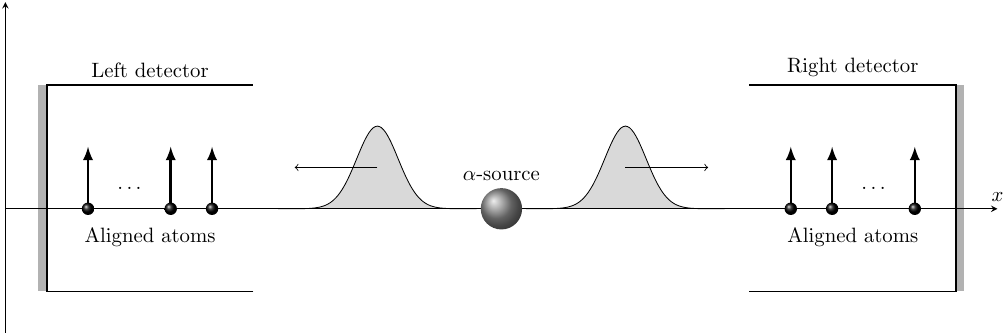}
\caption{Strongly simplified scheme of an $\alpha$-particle detection process in a cloud chamber containing a radioactive source.}
\label{model}
\end{figure}

This hypothesis was tested in reference \cite{sparenberg:13} and seemed to display a reduction of the spherical wave already through the interaction with a single atom. This explanation was however too simple: the observed effect is actually due to a misuse of the Born approximation, which is non unitary and cannot be used as such to evaluate absolute probability fluxes.
Here, to avoid this problem, we perform an exact calculation instead of using the Born approximation.
We also replace the complicated 3-dimensional problem by a simpler 1-dimensional version,
as recently proposed in reference \cite{carlone:15}.
There, the spherical wave is replaced by a linear superposition of two wave packets moving left and right
and the detector is schematically represented by two Dirac combs of equally-spaced ``atoms'',
symmetric with respect to the origin (see figure \ref{model}).
Each atom is described as a spin 1/2, which is in its down state initially and can be excited to its up ``ionized'' state at later times, through interaction with the $\alpha$ wave, hence leading to a possible detection.

Below we first introduce the model, discussing some of its features useful for numerical calculations.
Next we revisit the results of reference \cite{carlone:15} using a time-dependent approach,
limiting ourselves to a single spin on each side of the detector.
This provides a clear representation of the configuration-space structure of this system,
which explains the absence of multiple (left {\em and} right) measurements,
but also makes appear a limitation of the one-dimensional two-sided model.
Then, with the need to allow for arbitrary positions of the spins in mind,
we develop a time-independent approach.
There, only a one-sided model is considered, which removes the above-mentioned limitation.
We discuss the numerical results obtained up to 8 spins.
These reveal first hints that the spin positions could make measurements more or less likely.
We finish with some conclusions and perspectives for future works.

\section{The model}

In reference \cite{carlone:15}, it is proposed to describe the evolution of a quantum particle of mass $m$
(e.g.\ an $\alpha$ particle) in a quantum environment (a mesh of $N$ non-interacting non-moving spins 1/2)
within a deterministic model based on the Schrödinger equation.
The particle is thus described by a pure state, $\Psi(x, t)$, which is a spinor made up of $2^N$ components.
As already noticed by Mott \cite{mott:1929,mott:95} and as shown below, this complexity of the configuration space exponentially-increasing with the number of spins, i.e.\ with the size of the ``detector'', is the key element in the appearance of a classical-like behaviour for the quantum particle.

Each of the spinor components is a standard one-dimensional wave function $\psi_c(x,t)$, where $c=0, \dots, (2^N-1)$ represents a possible spin state of the mesh.
We found that a convenient way to index the spin states is to write $c$ in binary digits, 0 representing a down state and 1 representing an up state.
For instance, for a 6-spin case with 64 possible spin configurations, $c=53=110101_\mathrm{bin}$ represents a configuration where all spins are in the up state, except for the third and fifth ones.

The $\alpha$ particle only interacts with the spins at their locations $y_n$, $n=1, \dots, N$, where it can be either elastically scattered with a strength $\beta_n$ or inelastically scattered with a strength $\gamma_n$, loosing an energy $\epsilon_n$. Hence, in all generality the Hamiltonian reads (in units $\hbar^2/2m=1$)
\begin{equation}\label{H}
H = 
\bigoplus_{n=1}^N \left(\begin{array}{cc}
-\frac{\partial^2}{\partial x^2} + \beta_n\,\delta(x-y_n) + \epsilon_n & \gamma_n\,\delta(x-y_n) \\
\gamma_n\,\delta(x-y_n) & -\frac{\partial^2}{\partial x^2} + \beta_n\,\delta(x-y_n)
\end{array}\right)\:.
\end{equation}
In the simplest case, all spins being identical, one has $\gamma_n=\gamma, \beta_n=\beta, \epsilon_n=\epsilon$.

\begin{figure}[ht]
\centering
\includegraphics[scale=0.6]{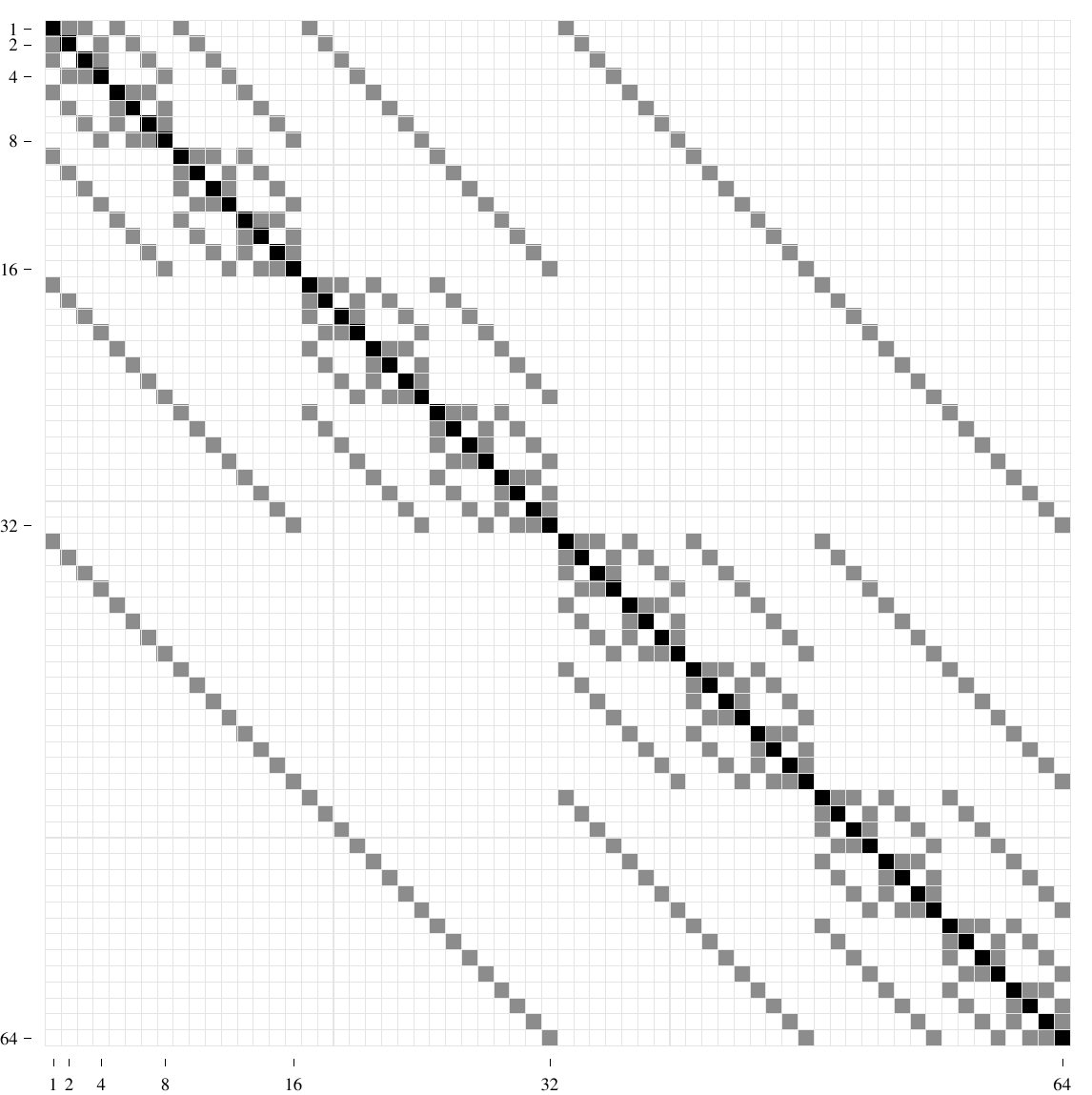}
\caption{The schematic sparse filling of the Hamiltonian $64\times 64$ matrix for six spins.}
\label{Hmat}
\end{figure}

Figure \ref{Hmat} schematically represents the structure of this Hamiltonian in the 6-spin case.
Due to the Dirac-peak interactions, each channel is only coupled to 5 other channels, which leads to a hollow fractal-like structure.

\section{Time-dependent approach and limitation of the 1D approach}

Let us first revisit the results obtained in reference \cite{carlone:15} using a time-dependent approach.
Space discretization is made using the same finite-difference scheme while time discretization is made using a time-exponential approach rather than a Crank-Nicolson scheme.

\begin{figure}[ht]
\centering
\includegraphics[scale=0.75]{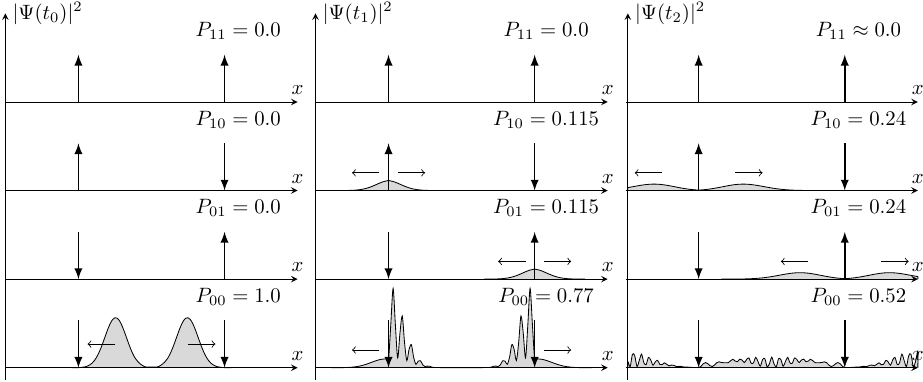}
\caption{Time evolution of a double gaussian wave packet emitted at the origin and propagating in a ``two-atom symmetric detector''.
The probabilities of finding the propagating particle in the $ij_\mathrm{bin}$ part of the state space is given by $P_{ij}$ at times $t_0<t_1<t_2$.
\label{timedep}}
\end{figure}

Figure \ref{timedep} represents the time evolution of a double gaussian wave packet, starting from the origin at time $t=0$ and evolving in a ``one-dimensional detector'' made up of two spin-1/2 atoms, symmetrically positioned on the left side and on the right side of the ``radioactive source''.
At time $t_0$, the wave packets have not reached the atoms yet and the state has only a non-vanishing component in the $00_\mathrm{bin}$ spin configuration.
At time $t_1$, the left-propagating wave packet interacts with the left atom, populating the $10_\mathrm{bin}$ component of the state.
Symmetrically, the right-propagating wave packet excites the right-hand-side atom, populating the $01_\mathrm{bin}$ component.
At time $t_2$, the excited configurations are populated with new gaussian wave packets, with a probablity of 24\% each.
The probability of non excitation is 52\%.
Remarkably, the probability of exciting both atoms together, which would correspond to a simultaneous detection on the left- and right-hand sides
is practically 0 at time $t_2$.
It is in this sense that the state-space structure explains the absence of ``left {\em and} right'' measurements,
as stressed in reference \cite{carlone:15}.
This is the one-dimensional version of Mott's original argument \cite{mott:1929,mott:95}
which states that in a 3-dimensional cloud chamber two atoms can be simultaneously excited only if they are aligned with the radioactive source.

In the present one-dimensional system, the situation is a bit less simple though, as guessed from figure \ref{timedep}:
at later times, both atoms will be excited, populating configuration $11_\mathrm{bin}$,
because of the reflected wave packets in configurations $10_\mathrm{bin}$ and $01_\mathrm{bin}$.
This is a limitation of the one-dimensional model, which could have been better emphasized and numerically explored in reference \cite{carlone:15}.
Such a reflected wave also appears in a 3-dimensional scattering situation but with an extremely small probability,
making it negligible:
the high energy of the incident particle leads to a very thin scattering cone
(typically $10^{-4}$ rad angular aperture for a 1-MeV $\alpha$-particle incident on a 1-\AA-radius atom).
In the following, to avoid this limitation of the one-dimensional model,
we only consider a one-sided detector.
This simplifies the problem as the number of spins is smaller;
moreover, the two-sided system can always be reconstructed by linear superposition of two one-sided systems.

\section{Time-independent approach}

Having switched to a one-sided detector, a time-independent approach can be used, meaning that the Schrödinger equation is solved for a given energy $E=k_0^2$ defined with respect to the threshold energy of the $0\dots0_\mathrm{bin}$ ground channel.
Time-dependent solutions can always be built by linear combination of these single-energy solutions,
which can be seen as a basis for time-dependent wave-packet construction.
A time-independent approach allows us to solve the problem analytically (for a very small number of spins),
symbolically (for a small number of spins) or numerically with algebraic methods (potentially for an arbitrary number of spins).
Here, we have chosen the analytic approach for a single spin, which allows us to define a so-called ``spin entropy''.
Next, we solve the problem numerically and provide results for up to 8 spins.
Symbolic calculations are deferred to a future work.

\subsection{One spin}

\begin{figure}[ht]
\centering
\includegraphics{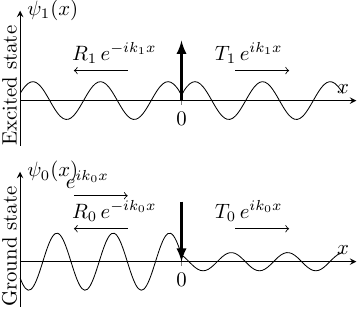}
\caption{Stationary solution corresponding to a plane wave incident on a single spin 1/2 in its ground state. \label{wf}}
\end{figure}

Let us start by solving analytically the single-spin case.
The solution of the stationary Schrödinger equation is expressed in terms of 4 coefficients defined in figure \ref{wf},
which allow us to calculate reflection and transmission probabilities in the elastic and inelastic channels.
The continuity conditions on the wave-function components lead to the following expressions for these coefficients,
obtained by solving a four-dimensional algebraic system,
\begin{equation}\label{eq:One_spin_solution}
\begin{cases}
R_1 = T_1 = \frac{2\mathrm{i} k_0 \gamma}{(\beta-2\mathrm{i} k_0)(\beta-2\mathrm{i} k_1) - \gamma^2}, \\
R_0 = \frac{2\mathrm{i} k_1 \beta - \beta^2 + \gamma^2}{(\beta-2\mathrm{i} k_0)(\beta-2\mathrm{i} k_1) - \gamma^2}, \quad
T_0 = \frac{-2\mathrm{i} k_0 (\beta-2\mathrm{i} k_1)}{(\beta-2\mathrm{i} k_0)(\beta-2\mathrm{i} k_1) - \gamma^2} \:,
\end{cases}
\end{equation}
where $k_1=\sqrt{k_0^2-\epsilon}$ is the wave number in the excited channel.
Figure \ref{wf} also presents a typical stationary solution.
The continuity of the solution, together with the discontinuity of its first derivative due to the Dirac-peak interaction,
can be seen.

Such solutions can be advantageously characterized by a ``spin entropy'', summarizing the splitting of the solution among the different spin states. For the single-spin case, this entropy reads
\begin{equation}
 S = -P_\mathrm{exc} \ln P_\mathrm{exc} - P_\mathrm{gnd} \ln P_\mathrm{gnd},
\end{equation}
where the excitation probability $P_\mathrm{exc}$ reads
\begin{equation}
 P_\mathrm{exc} = \frac{k_1}{k_0} \left(|R_1|^2+|T_1|^2 \right)
 = \frac{8 k_0 k_1 \gamma^2}{4\beta^2(k_0+k_1)^2 + (4 k_0 k_1-\beta^2 +\gamma^2)^2}
\end{equation}
and the ground-state probability $P_\mathrm{gnd} = |R_0|^2+|T_0|^2 = 1-P_\mathrm{exc}$.

\begin{figure}
\centering
\includegraphics{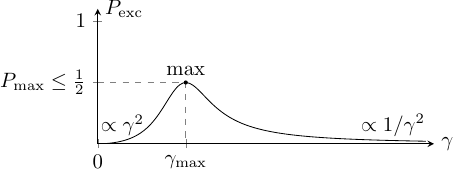}
\caption{Excitation probability as a function of the coupling strength $\gamma$ for the single-spin case. \label{Pexc}}
\end{figure}

The most interesting dependence of this excitation probability is on the coupling strength $\gamma$,
as seen in figure \ref{Pexc};
it displays a maximum value for $\gamma=\gamma_\mathrm{max}$
which can be interpreted as the optimal coupling strength for particle detection.
When $\epsilon=\beta=0$, the excitation probability simplifies to
\begin{equation}
 P_\mathrm{exc} = |R_1|^2+|T_1|^2 = \frac{8 k_0^2 \gamma^2}{(4 k_0^2+\gamma^2)^2} \quad (\epsilon=\beta=0),
\end{equation}
which leads to a maximal spin entropy of ln 2 for $\gamma_\mathrm{max}=2k_0$.
In the single-spin case, the maximum detection probability is thus 50\%.
For other values of $\epsilon$ and $\beta$, this probability is decreased but the general trend of the $\gamma$ dependence remains the same.

\subsection{$N$ spins}

Let us finally proceed to the general case of $N$ spins.
As a first approach we developed a purely numerical algorithm based on the construction and resolution of a $2N 2^{N}$-dimensional linear-algebra system obtained by combining the continuity equations satisfied by all the wave-function coefficients
$R_{ci}, T_{ci}$ with $c\in[0,2^N-1]$, $i\in[1, N]$.
No particular attempt was made to exploit the hollow structure of the Hamiltonian shown in figure \ref{Hmat}.
Hence, the algorithm is limited to $N<9$ on a standard personal computer.
In future works, we plan to improve this algorithm and to solve the problem both symbolically and numerically.

These preliminary results already prove very interesting.
To discuss them, we introduce the $\bw$-spin excitation probabilities $P(\bw)$,
where $\bw$ is the Hamming weight, i.e.\ the number of 1-bits in a binary sequence representing an integer.
Following reference \cite{carlone:15},
we define the ground-state, one-spin excitation and track-detection probabilities in terms of these excitation probabilities as
\begin{equation}
\underbrace{P(\bw=0)}_{P_\mathrm{gnd}} +
\underbrace{P(\bw=1)}_{P_\mathrm{OS}} +
\underbrace{P(\bw\geq 2)}_{P_\mathrm{trk}} = 1.
\end{equation}
Table \ref{spin-exc} shows these probabilities for a regularly-spaced $N$-spin network.
Their general behavior can be interpreted through an entropic effect:
the no-spin excitation probability is close to 50\% for small $N$ and regularly decreases with increasing $N$,
as it becomes a more and more exceptional configuration among the $2^N$ possible configurations.
This shows that, contrary to the single-spin case, the probability of no excitation in the detector can become much smaller than 50\%. 
The one-spin excitation (second line of table \ref{spin-exc}), corresponding to $N$ possible configurations,
stays roughly constant.
In contrast, the track-detection probability,
corresponding to $2^N-N-1$ possible configurations (third line of table \ref{spin-exc}),
regularly increases,
which is a key feature of the detection mechanism.
\begin{table}
\caption{\label{spin-exc}
Probabilities for no-spin ($P_\mathrm{gnd}$), one-spin ($P_\mathrm{OS}$) or more-than-one-spin ($P_\mathrm{trk}$) excitations
for a particle of energy $\pi^2$ incident on a regularly-spaced $N$-spin network of step 0.1
for $\epsilon=0.01$, $\beta=0.5$ and $\gamma=3$.}
\begin{tabular}{c|cccc}
\hline\noalign{\smallskip}
$k_0=\pi, \gamma= 3 $          & $N=2$     &  $N=4$      & $N=6$       & $N=8$       \\
\noalign{\smallskip}\hline\noalign{\smallskip}
$P_\mathrm{gnd}$  & $0.60514$  & $0.41826$    & $0.24661$      & $0.15813$   \\
$P_\mathrm{OS}$ & $0.28992$  & $0.36631$    & $0.39652$      & $0.37044$   \\
$P_\mathrm{trk}$   & $0.10494$  & $0.21543$   & $0.35687$     & $0.47143$ \\
\noalign{\smallskip}\hline
\end{tabular}
\end{table}

Table \ref{Ptrk} confirms these results but also shows that these simple trends are not universal.
There it is seen that the track-detection probability can strongly vary for some specific numbers of spins.
This seems to be the case in particular for large energies and
when the single-spin excitation is maximal, i.e.\ for $\gamma=\gamma_\mathrm{max}$
(see the surprisingly small value obtained for $k_0=400/3$, $\gamma=800/3$, $N=4$).
This suggests that, at least in the present simplified one-dimensional model,
a track detection could be seen as an ``inverse-resonance'' phenomenon.
``Detection'', i.e.\ excitation of a large number of spins,
may indeed depend on the precise match of the number and of the position of the atoms with the particle energy.

For a two-sided detector,
detection would then only occur on the side where such a matching condition is satisfied by the atomic configuration,
whereas the wave packet would be reflected on the other side and hence would not excite a sufficient number of atoms to be detected.
Though the initial particle wave packet would be symmetric, with 50 \% of the wave going in each direction
(cf first panel of figure \ref{timedep}),
the microscopic state of the detector would break that symmetry,
hence leading to an enhanced detection probability on one side (the side for which the ``inverse-resonance'' condition is best satisfied).
Moreover, all present calculations are performed assuming an initial ground state for the detector, i.e.\ a zero-temperature state
(all spins down at $t=0$ in figure \ref{timedep});
an additional way of breaking symmetry for the detector initial state
would be to consider an excited state with random phases mimicking a non-zero-temperature thermal state.

Going back to a 3-dimensional situation, detection would then also occur in the direction where the configuration of the detector atoms
best match the particle energy.
For a given particle energy, such a direction would always come up, due to thermal randomization of the detector configuration:
the random positions of the atoms would lead to a quasi infinite set of different scattering parameters at the same time,
depending on the direction.
\begin{table}
\caption{Probabilities of ``track detection'', i.e.\ more-than-one-spin excitation,
for a particle of energy $k_0^2$ incident on a regularly-spaced $N$-spin network of step $0.05/N$
for $\epsilon=0.04$, $\beta=10^{-4}$ and different values of $\gamma$. \label{Ptrk}}
\begin{tabular}{c|cccc}
\hline\noalign{\smallskip}
 $P_\mathrm{trk}, k_0=400/3$ & $\gamma = 50$ & $\gamma = 100$ & $\gamma = 150$ & $\gamma_{\max} = 800/3$ \\
\noalign{\smallskip}\hline\noalign{\smallskip}
$N = 2$ & $0.0074691$   & $0.063126$     & $0.14733$      & $0.24907$            \\
$N = 3$ & $0.015572$    & $0.12502$      & $0.21374$      & $0.2812$              \\
$N = 4$ & $0.032093$    & $0.25924$      & $0.39707$      & $0.066758$           \\
$N = 6$ & $0.07459$     & $0.40374$      & $0.51122$      & $0.22914$             \\
$N = 7$ & $0.099126$    & $0.47012$      & $0.40409$      & $0.37756$             \\
$N = 8$ & $0.12356$     & $0.49174$      & $0.50721$      & $0.32307$ \\
\noalign{\smallskip}\hline
\end{tabular}
\end{table}

\section{Conclusions and perspectives}

In conclusion, the present paper revisits Mott's seminal measurement problem \cite{mott:1929,mott:95} following the one-dimensional model proposed in reference \cite{carlone:15}.
Our approach differs from this last reference in two respects:
first, our aim is more ambitious as we aim at answering the question of why a particular track is chosen instead of just explaining why no multiple track is observed.
For that, we develop a model allowing us to arbitrarily choose the atom positions,
as it is envisaged that these positions could be the apparatus hidden variables determining the measurement outcome.
Second, to avoid artificial reflections due to the one-dimensional character,
we limit ourselves to a one-sided detector.
This divides the number of spins by two and allows us to use a time-independent approach which is more promising for large-scale calculations.

In the single-spin case, an analytical result is obtained, which shows that the most important parameter of the model is $\gamma$,
the coupling strength between the different spin states.
The other parameters, $\epsilon$ and $\beta$, which represent the energy loss during the spin excitation and the elastic scattering on the spin,
can probably be chosen to zero without much loss in the physical content of the model.
This conclusion should hold for the $N$-spin case, except if one intends to model the energy loss of the particle in the detector,
e.g.\ to calculate the detector stopping power.

For the $N$-spin case, first numerical results are obtained for an equally-spaced mesh.
They are limited to 8 spins for computational-cost reasons but are however sufficient to reveal that track-formation probabilities indeed depend on the detector microscopic state
(the number of atoms in the present case).
We suspect this behavior to be due to a kind of ``inverse-resonance'' phenomenon,
in the sense that for a given energy of the incident wave, the transmission through the spin network
would only be possible for particular positions of the detector atoms.
The atoms would reach this condition randomly, due to thermal agitation.
This is thus the opposite of the traditional resonance phenomenon,
which occurs when the wave energy matches some complex energy eigenvalues \cite{barra:99}.

In the near future, we plan to further study this phenomenon, combining symbolic and numeric calculations,
and to extend our study to randomly-spaced spin meshes.
In that context, we expect that randomization may lead to an Anderson localization \cite{anderson:1958},
which should be particularly strong in a one-dimensional model.
We expect this effect to be even stronger in the presence of spin excitations,
which make the structure of the state space even richer and more complex,
hence allowing for a spreading of the wave over the entire spin configuration space.
With such a generalized Anderson localization,
a simple solution to the measurement problem in one dimension would then be that the wave is generally reflected on the spin network 
and is only transmitted (and measured) when the atom configuration matches the wave energy.
Another effect which could be studied in this one-dimensional model is the impact of the initial thermal state of the spins,
which would then stand for the internal degrees of excitation of the detector atoms or molecules.
These could also be seen as microscopic variables determining the measurement result.

On a longer term, a full 3-dimensional quantum model should be set up (see also the semiclassical approach of reference \cite{dellantonio:15}).
Spin-dependent point interactions could be used as a first step \cite{cacciapuoti:07,figari:14,figari:16}, before going to more realistic particle-atom interactions.
Such models would allow us to get free of the strong constraints due to the one-dimensional model,
in particular regarding Anderson localization.
It would be interesting to test to which extent the localization occurring in one-dimension holds in three dimensions,
given the expected state-space contribution to this localization.

Developing such a model with realistic orders of magnitudes for elastic-scattering and ionization probabilities could be very useful
not only in the context of the quantum-measurement problem in a cloud-chamber apparatus.
It could provide a microscopic approach to existing macroscopic effective and non-linear models,
like the stopping-power formula by Bethe \cite{bethe:1930},
and replace these models when they are less reliable, in particular in low-energy situations. 
On an even longer term, combining these microscopic and macroscopic models could be applied to the development of new track detectors,
like the active targets used in nuclear physics \cite{beceiro-novo:15} or the huge CLOUD cloud chamber at CERN \cite{kirkby:16},
which is dedicated to the study of cloud formation due to high-energy radiation present in the high atmosphere.

Regarding the measurement problem, our approach is similar and complementary to the very detailed detector description performed in reference \cite{allahverdyan:13},
where the relevant microscopic degrees of freedom could be explicitly taken into account.
There, the full thermodynamical process taking place during a measurement could be tackled, in addition to the preliminary switch from linear superposition to statistical mixture.
We hope a similar sophistication could be reached someday for track detectors too.
We suspect tracing out the microscopic degrees of freedom, i.e.\ the spin states,
could lead to the increase of the ``spin entropy'' for the scattered wave.
We conjecture the measured track could correspond to a minimal spin-entropy increase,
as the corresponding wave would be far reaching, with a higher probability of leading to a track formation.
Other waves would be stopped by the Anderson-like localization mentioned above,
due to a spreading of the wave over the spin configuration space.
  Note that the entropy of the total system, that would take into account the interaction between the atoms and the full vapor-condensation process,
would still be maximal, contrary to the spin entropy, since the track formation corresponds to a phase change with a strong entropy increase.

Let us finally stress that, once again, our philosophical starting point is more ambitious (or just more naive!) than the ones adopted
in references \cite{carlone:15} or \cite{allahverdyan:13} or in decoherence theory,
where no attempt is made at finding a deterministic explanation for a particular measurement outcome.
We consider that as long as this possibility has not been definitely ruled out,
it is an option worth exploring as it could indeed bring quantum mechanics back into the world of deterministic science.

\begin{acknowledgements}
This text presents research results of the IAP program P7/12 initiated by the Belgian-state Federal Services for Scientific, Technical, and Cultural Affairs.
We thank Ruben Ceulemans (KU Leuven) for useful discussions and for the check of our numerical results.
\end{acknowledgements}


\end{document}